\documentclass[12pt]{article}

\usepackage{amsfonts}
\usepackage{graphicx}
\def\QTR#1#2{{\csname#1\endcsname #2}}
\def\Bc{{\cal B}\,}

\def\Ah{\hat{A}\,}
\def\Qh{\hat{Q}\,}

\def\Bh{\hat{B}\,}

\def\Ib{\mathbb{I}\,}

\def\eopp{{ \vrule height7pt width7pt depth0pt} }
\def\Prf{\noindent {\em Proof:}\, }

\makeatletter
\@addtoreset{equation}{section}
\makeatother
\newtheorem{thm}{Theorem}[section]
\newtheorem{defn}[thm]{Definition} 
\newtheorem{lem}[thm]{Lemma} 
\newtheorem{cor}[thm]{Corollary} 
\newcommand{\BEQ}{\begin{equation}}
\newcommand{\EEQ}{\end{equation}}
\newcommand{\NEQ}{\end{equation}}
\newcommand{\BEC}{\begin{center}}
\newcommand{\NEC}{\end{center}}

\def\eopp{{ \vrule height7pt width7pt depth0pt} }
\def\Prf{\noindent {\em Proof:}\, }

\begin{document}

\bibliographystyle{plain}
\title{Banded Matrix Fraction Representation of Triangular Input Normal Pairs}

\author{
Andrew P.~Mullhaupt$^*$ and Kurt S.~Riedel${^\dagger}$
\thanks{S.A.C. Capital Management, LLC,
540 Madison Ave, New York, NY 10022, \ \
$\dagger$ 
Present Address: Millennium Partners, 666 Fifth Ave, New York 10103}
} 

\maketitle

\begin{abstract}
An input pair $(A,B)$ is triangular input normal if and only if
$A$ is triangular and $AA^* + BB^* = \Ib_n$, where $\Ib_n$ is the
identity matrix. Input normal pairs generate an orthonormal basis for the
impulse response.
Every input pair may be transformed to a triangular input normal pair.
A new system representation is given: $(A,B)$ is triangular normal
and $A$ is a matrix fraction, $A=M^{-1}N$,
where $M$ and $N$ are triangular matrices
of  low bandwidth. For single input pairs, $M$ and $N$ are bidiagonal
and an explicit parameterization is given in terms of
the eigenvalues of $A$.
This band fraction structure allows for
fast updates of state space systems and fast system identification.
When $A$ has only real eigenvalues, one state advance requires
 $3n$ multiplications for the single input case.
\end{abstract}

EDICS Numbers: 2-SREP, 2-SYSM


\BEC
\section{INTRODUCTION.} \label{I}
\NEC 

An increasingly popular technique in system identification is
the use of orthonormal forms with fixed denominators
\cite{MRbook,Wahlberg,HVB,NG,VHB,NHG,Ninness}. In this approach,
the unknown impulse response, $G(q;\{c_k\})$, is represented as
a linear sum of orthonormal basis functions:
\BEQ \label{IIRexp}
G(q,\{c_k\}) = \sum_{k=1}^{n} c_k \Bc_k(q) \ ,
\NEQ
where $q$ is the forward shift operator. Here $\Bc_k(q)$ are prescribed
rational transfer functions with poles $\{ \lambda_j, j=1\ldots n\}$.
The unknown coefficients are then identified using least squares
identification or a robust analog \cite{LS}.
The utility of this approach is now widely documented \cite{Ninness,NHG}.
In system identification, the use of orthonormal pairs
improves the condition of the estimate \cite{Ninness,MR4}.
We specialize to the case where $\Bc_k(q)$ corresponds to the $k$th row
in the semi-infinite impulse response  matrix, $\Omega(A,B)$:
\BEQ
\Omega(A,B) \equiv (B, AB, A^2B \ldots ) \ .
\NEQ
Here $A$ is the $n\times n$  state transition matrix
and $B$ is a $n \times d$  matrix with $n \ge d$.
 Let $C$ be the $p\times n$ matrix whose $k$th column is $c_k$. The $j$th
term or lead of the impulse response in (\ref{IIRexp}) is $CA^{j-1}B$.

In this paper, we develop a computationally convenient representation
of the state transition matrices that arise from orthonormal bases. We then
show how these representations allow the rapid identification of
orthonormal single input models with prescribed poles.
Our system representation results are general and well-suited for filter
design as well. In the single input case, our filters are a particular
realization of the orthogonal filters of \cite{HVB,NG}.

In system representation theory, one seeks families of
system triples, $(A,B,C)$,
that parameterize the set of all transfer functions subject to the similarity
equivalence.
(Two systems are equivalent if there exists an invertible matrix, $T$,
such that  $(A',B',C')$ $= (TAT^{-1},TB,CT^{-1})$). We examine
representations of the input pair, $(A,B)$.
The input pair is in triangular input normal (TIN) form if and only if
 $AA^* +BB^*  = \Ib_n$,
with $A$  {\em lower triangular} (LT) and  $\Ib_n$ is the identity matrix.
The input normal condition is equivalent to the row orthogonality of
the semi-infinite impulse response matrix, $\Omega$.


The state transition matrix $A$ is represented as a matrix fraction:
$A=M^{-1}N$. Our results are not related to the extensive literature
on representations of the transfer function $G(q)$ as a matrix fraction:
 $G(q)= N(q)/D(q)$.
We show that generic linear state space systems with 
forcing $\epsilon_t$ have
an equivalent representation of the form:
\BEQ \label{MNSAdv}
M z_{t+1} = N z_t + \Bh \epsilon_t \ ,
\NEQ
where $M$ and $N$ are lower triangular matrices with bandwidth $d$.
In addition,  $M$ and $N$ are chosen such that
{\em the covariance matrix of the state vector $z_t$
tends to the identity matrix} for white noise forcing.

We believe that our work is the first systematic study of system representations,
where $A$ is a matrix fraction of banded matrices, $A=M^{-1}N$.
The band fraction, $A=M^{-1}N$, is never computed in practice.
To advance the state vector,
we first compute $r_t \equiv N z_t + \Bh \epsilon_t$ and then solve
the banded matrix equation $M z_{t+1} =r_t$. Constructing $r_t$ requires
 $n(d+1)$ multiplications and solving for $z_{t+1}$
requires $nd$ multiplications since $M_{i,i}=1$. Thus the total multiplication count 
for a state vector update is $(2d+1)n$. 
For SISO systems with real eigenvalues, the state transition matrix
is identical to the cascade systems proposed in \cite{HVB}.
However, the parameterization/realizations of the $(A,B)$ pair differ.



In Section \ref{TINSect}, the existence and identifiability of TIN pairs
is examined. Generically, TIN pairs are unique up to reordering the
eigenvalues and
phase rotations of the coordinates:
 $ B_{j,k}\leftarrow B_{j,k} \exp i(\theta_j)$,
 $ A_{i,j}\leftarrow A_{i,j} \exp i(\theta_i-\theta_j)$.


In Sections \ref{BandFracSect}-\ref{BandFracSectM},
we prove that generic TIN pairs have
a band fraction representation: $A=M^{-1}N$, $B=M^{-1}\tilde{B}$,
where $M$ and $N$ are triangular matrices of bandwidth $d$.
We derive the eigenvectors of $A$ in the $d=1$ case
and prove the numerical stability of the representation
when the eigenvalues of $A$ are order in increasing magnitude.
In Section \ref{NumSect}, we discuss the numerical details of
the band fraction representation and other realizations of IN filters.
We relate our band fraction representation in the $d=1$ case,
to the orthonormal basis functions of \cite{NG}.
In Section \ref{SysIdentSect}, we show the advantage of our representations
in system identification.

{\em Notation:}
In the remainder of the article,
we suppress the word `lower''. (`Triangular'' means lower triangular.)
Here $A$ is a $n\times n$ matrix
with eigenvalues $\{\lambda_i\}$
By $A_{i:j,k:m}$, we denote the $(j-i+1)\times (m-k+1)$ subblock
of $A$ from row $i$ to row $j$ and from column $k$ to column $m$.
We abbreviate $A_{i:j,1:n}$ by $A_{i:j,:}$.
The matrix $A$ has upper bandwidth $d$ if $A_{i,j}=0$ when $j> i+d$.
We denote by $(B | A)$ the $n \times (n+d)$ matrix formed by the
concatenation of $B$ and $A$.
`Stable'' means $|\lambda|<1$.
The $n \times n$ identity matrix is $\Ib_n$ and
$e_k$ is the unit vector in the $k$th coordinate.
We define the equivalence $(A,B) \approx (A_2,B_2)$
when $({ A_2} \equiv { EAE}^{-1}, { B_2}\equiv { EB})$, where
 $E_n$ is a diagonal unitary matrix: $E_{j,k}=\exp(i\theta_j)\delta_{j-k}$.
\section{TRIANGULAR INPUT NORMAL PAIRS}
 \label{TINSect}

In this section, we present the fundamental representation results
for the reduction to triangular normal form.

\begin{defn} \label{DefTIN}
An input pair, $({ A},{ B})$, is
input normal  (IN) 
if and only if
\BEQ \label{INeq}
{ \Ib_n} -{ A} { A}^* =   { B} {B}^* \ \ \ .
\EEQ
The input pair is a TIN pair if it is input normal
and $A$ is lower triangular.
\end{defn}

IN pairs are not required to be stable or controllable.
(From (\ref{INeq}), $A$ must be at least marginally stable.)
In \cite{Moore},  `input balanced'' has a more restrictive
definition of (\ref{INeq}) and the additional requirement that the
observability Grammian be diagonal. We do not impose any such condition
on the observability Grammian.
We choose this language so that `normal'' denotes restrictions on only
one Grammian while `balanced'' denotes simultaneous restrictions on
both Grammians.

Thus $(A,B)$ is an IN pair if and only if the concatenated $n \times (n+d)$
matrix $(A | B)$ is row orthogonal.
When $(A,B)$ is input normal, then the identity matrix solves Stein's equation
(also known as the discrete Lyapunov equation):
\BEQ \label{SteinEq}
P -{ A}P { A}^* =   { B} {B}^* \ \ \ .
\EEQ
For stable $A$, $P = \sum_{j=0}^{\infty} A^{j} B{B}^*  A^{j*}$,
So $(A,B)$ is input normal when  $\Omega(A,B)^*\Omega(A,B)$ $ =\Ib_n$ or
equivalently when the basis, $\{\Bc_k(q)\}$,
constitutes an orthonormal set.

\begin{lem} \label{QUniq}
Let $(A,B)$ and  $(A',B')= (TAT^{-1},TB)$ be equivalent IN pairs,
with $(A,B)$ stable and controllable. Then $T$ is unitary.
\end{lem}

\Prf
Both $\Ib_n$ and $T^{-1}T^{-*}$ solve the Stein equation. Since
 $(A,B)$ is stable and controllable,
the solution of Stein's equation is unique.
\eopp


\begin{thm}
\label{Mthm}
Every  stable,  controllable input pair
$(A, B)$, is similar
to a  lower
triangular  input normal pair
$(\tilde{ A} \equiv { TAT}^{-1}, \tilde{ B}\equiv { TB})$
with $\|\tilde{B}\|^2 \le 1$.
The order of the eigenvalues of $\tilde{ A}$ may be specified arbitrarily.
If $(A,B)$ is real and $A$ has real eigenvalues,
then $(\tilde{ A}, \tilde{ B})$ and $T$
may be chosen to be real.
\end{thm}

\Prf
Let $L$ be the  unique Cholesky lower triangular factor of
${ P}$
with positive diagonal entries:
${ P} ={ LL}^{*}$.
Here ${ L}$ is invertible. 
We set $\hat{ A}={ L}^{-1}{ AL}$ and
$\hat{ B}={ L}^{-1}{ B}$.
Note $(\hat{ A}, \hat{ B})$ satisfies (\ref{INeq}).
By Schur's unitary triangularization theorem (Horn and Johnson 2.3.1),
there exists a unitary matrix, ${ Q}$ such that
$Q  \hat{ A}{ Q}^*$ is lower   triangular.
The proof of the real Schur form is in Sec.\ 7.4.1 of \cite{GVL}
and Theorem 2.3.4 of \cite{HJ}.
The eigenvalues of $\tilde{ A}$
may be placed in any order \cite{HJ}.
Clearly, $\tilde{ A} = { Q}  \hat{ A}{ Q}^*$ and $\tilde{B} = { Q}\hat{ B}$
satisfy (\ref{INeq}) and therefore $\|\tilde{B}\|^2 \le 1$.
\eopp

We now study the number of equivalent TIN pairs:

\begin{thm} \label{SchurUniq0}
Let $A$ and $\Ah$ be $n\times n$ LT matrices
with $A_{ii}=\Ah_{ii}$ and let $A$ and $\Ah$ be unitarily
similar: $\Ah U =U{A}$ with $U$ unitary.
Let $m$ be the number of distinct eigenvalues. Partition
$A$, $\Ah$ and $U$ into $m$ blocks corresponding to the
repeated eigenvalue blocks. Let $n_i$ be the multiplicity of the $i$th
eigenvalue, $1\le i \le m$. Then $U$ has block diagonal form:
 $U = U_1 \oplus U_2 \oplus \ldots \oplus U_m$, where $U_i$ is a unitary
matrix of size $n_i \times n_i$.
\end{thm}

\Prf
From
$ \hat{A}_{m, m} U_{m,1} =  U_{m,1} A_{1,1}$.
If $m>1$, then $ \hat{A}_{m, m}$ and $ A_{1,1}$ have no
common eigenvalues. By Lemma 7.1.5 of \cite{GVL}, $U_{m,1}\equiv 0$.
Repeating this argument shows $U_{m-k,1}\equiv 0$
 for $k=0,1\ldots <m-1$.
By orthogonality, $U_{1,j}= 0$ for $1<j<m$.
We continue this chain
showing that  $U_{i,2} = 0$ for $i \ne 2$, etc.
Proof by finite induction.
\eopp

\begin{cor} \label{SchurUniq}\label{SchurUniqCor}
Let $A$ be an $n\times n$ matrix with distinct eigenvalues.
Then $A$ is unitarily similar to triangular matrix $\hat{A}$
with ordered eigenvalues and $\hat{A}$ is unique up to diagonal
unitary similarities:
 $\hat{A} \leftarrow E \hat{A}E^*$,
where $E_{i,j}=\exp(i \theta_j)\delta_{i-j}$.
\end{cor}

If $A$ is invertible and $(A,B)$ is TIN,
then $A$ is the Cholesky factor
of $\Ib_n -BB^*$ times a diagonal unitary matrix:

\begin{thm}\label{AlterSchur} \label{CholFact}
Let $(A,B)$ be a TIN pair with rank $\left(\Ib_n - BB^* \right)=n$,
then
\BEQ
A = {\rm Cholesky}\left(\Ib_n - BB^* \right) E_{n} \ ,
\NEQ
where $ E_n$ is a diagonal unitary matrix.
In the real case, $E_n$ is a signature matrix.
\end{thm}

\Prf
Let $L$ be the Cholesky factor of $\left(\Ib_n - BB^* \right)$.
Clearly $A= LQ$ some unitary $Q$ with $L_{ii} \ge 0$.
By the uniqueness of the LQ factorization,
$L$ and $Q$ are unique.
Note $Q=L^{-1}A$ is lower triangular and therefore
diagonal.
\eopp

\BEC
\section{BAND FRACTION REPRESENTATIONS OF SINGLE INPUT TIN PAIRS.}
\label{BandFracSect}
\NEC

For the single input case ($d=1$), TIN pairs have an explicit
band fraction representation $A=M^{-1}N$, $B=\rho_1M^{-1}e_1$, where
 $M$ and $N$ are bidiagonal.
The representation is parameterized by the eigenvalues of $A$.
For each eigenvalue, $\lambda_k$, of $A$, we define
 $\rho_k= \sqrt{1- | \lambda _{k}|^2}$,
 $\mu_k= \rho_{k+1} /\rho_k$ and $\gamma_k = \lambda_{k}^* \mu_k $.
We define the matrices:
\begin{equation} \label{Mdef} \label{Mbidiag}
M\equiv {\rm bidiag}\left(
\begin{array}{llllllll}
1 &  & 1 &  & 1 &  & \cdots  &  \\
 & \gamma_{1} &  & \gamma_{2} &  & \gamma_{3} &
 & \cdots \end{array} \right) \ ,
\end{equation}
\begin{equation}\label{Nbidiag} \label{Ndef}
N \equiv{\rm bidiag}\left(
\begin{array}{llllllll}
\lambda_{1} &  & \lambda_{2} &  &  \lambda_{3} &  & \cdots  &  \\
& \mu_{1} &  & \mu_{2} &  & \mu_{3} &  & \cdots
\end{array}
\right)  \ \ ,
\end{equation}
where the top row contains the diagonal elements of $M$ or $N$
and the bottom row contains the $(n-1)$ elements of first subdiagonal
of the bidiagonal matrices, $M$ and $N$.

\begin{thm}\label{BiDiagCon}
Let $\hat{A} \equiv M^{-1}N$ and $\hat{B}=M^{-1} \rho_1 e_1$,
where $M$ and $N$ are given by (\ref{Mbidiag}- \ref{Nbidiag})
with $|\lambda_k|<1$. Then $(\hat{A},\hat{B})$ is TIN with
eigenvalues $\{\lambda_k\}$.
\end{thm}

\Prf
Explicit evaluation shows $MM^* -NN^* = \rho_1^2 e_1e_1^*$.
\eopp

The Hautus criterion states that if a stable matrix, $A$,
is nonderogatory (There is only one Jordan block for each eigenvalue of $A$
in the eigendecomposition of $A$.), then there is a vector $B$ such that
$(A,B)$ is controllable. Thus the Hautus criterion together with
Theorem \ref{Mthm} imply that

\begin{cor}\label{BiDiagHautus}
Let $A$ be a stable, nonderogatory matrix. Then there exists a similarity
transformation, $T$, such that $\hat{A}=TAT^{-1}$ has the bidiagonal fraction
representation of Theorem \ref{BiDiagCon}.
\end{cor}

The bidiagonal fraction can be casted in a more aestetic form
$A=M^{-1}N=M_0^{-1}N_0$, where $M_0$ and $N_0$ are defined as
\begin{equation} \label{M0def}
M_0\equiv {\rm bidiag}\left(
\begin{array}{llllllll}
c_{1} &  & c_{2} &  & c_{3} &  & \cdots  &  \\
& s_{1}^{*} &  & s_{2}^{*} &  & s_{3}^{*} &  & \cdots
\end{array} \right) \ ,
\end{equation}
\begin{equation} \label{N0def}
N_0\equiv{\rm bidiag}\left(
\begin{array}{llllllll}
s_{1} &  & s_{2} &  & s_{3} &  & \cdots  &  \\
& c_{1} &  & c_{2} &  & c_{3} &  & \cdots
\end{array}
\right)  \ \ ,
\end{equation}
with $c_{k}={1}\left/\sqrt{1-\left| \lambda _{k}\right|^2} \right.$ and
$s_k \equiv \left. \lambda_k\right/\sqrt{1-| \lambda _{k}|^2} $.

In a series of excellent articles
\cite{Wahlberg, HVB,NG,VHB,NHG,Ninness},
a number of researchers have
constructed IN filters using cascade realizations.
We now consider the case single input case where $A$ has real
eigenvalues. In this case, the cascade realization of the IN
filter is triangular. By Corollary \ref{SchurUniqCor}, there is
a unique triangular state advance matrix with the given eigenvalue
ordering (up to diagonal similarity transformations with $\pm 1$
elements).
Thus all state space constructions of IN filters are simply different
realizations of the same $(A,B)$ pair (up to sign flips of the coordinates).
In Section \ref{NumSect}, we discuss the numerics of the various realizations
of IN  filters.

We now give a more constructive proof of equivalence of the
band fraction representation with the orthogonal basis functions
of \cite{NG}.  
Given a set of decay rates/poles of the frequency responses,
$\{ \lambda_n \}$, Ninness and Gustafsson derive a set of orthonormal basis
functions in the frequency domain:
\BEQ
\hat{z}_{n}\left( w\right)\ = \ \sum_{t=0}^{\infty }z_{n}\left( t\right) w^{-t}
\ =\
w^{\alpha}\frac{\sqrt{1-\left| \lambda _{n}\right| ^{2}}}{w-\lambda _{n}}%
\prod_{k=1}^{n-1}\left( \frac{1-\lambda _{k}^{*}w}{w-\lambda _{k}}\right) ,
\NEQ 
where $\alpha$ is $0$ or $1$. From this, we derive the relation:
\begin{equation}
({w-\lambda _{k}})\hat{z}_{k}\left( w\right)/{\sqrt{1-\left| \lambda _{k}\right| ^{2}}}
=({1-\lambda _{k-1}^{*}w})
\hat{z}_{k-1}\left( w\right)/{\sqrt{1-\left| \lambda_{k-1}\right| ^{2}}} \ \ .
\end{equation}
for $k\ge 1$ with $\lambda_0\equiv 0$ and $\hat{z}_0(w) \equiv z^{\alpha}$.
Equating like powers of $w$ yields
\begin{equation} \label{IN update}
z_{k}\left( t+1\right) + \lambda_{k-1}^{*} \mu_{k-1}z_{k-1}\left( t+1\right)
=\lambda_{k}z_{k}\left( t\right) + \mu_{k-1}z_{k-1}\left( t\right) \ ,
\end{equation}
with $z_0(t) \equiv \delta_{0,-\alpha}$.
This shows that the bidiagonal matrix fraction representation
generates the same basis functions as in \cite{NG}.
The $k$-th component of $z_t$  is to be identified with the $t$-th lead of
 the $k$-th orthonormal transfer function, ${\cal B}_k(q)$.



We show that if the eigenvalues are sorted
in order of increasing magnitude, then $M^{-1}$
does not have large elements.
We now give two results concerning the bidiagonal
matrix fraction representation $(d=1)$. The first result shows that if the
eigenvalues are sorted in order of increasing magnitude,
then the bidiagonal factorization
is well-conditioned. We then give the eigenvectors of $A$.

{\em A. Explicit $M$ inversion.}

We can interpret the bidiagonal matrix fraction representation
as an LR factorization of the augmented system:
$(B | A)= M^{-1} (e_1| N)$.
We now explicitly invert $M$ in (\ref{Mdef}) and show that the
numerical conditioning of the matrix inversion is good without pivoting.
We define the $x-$condition number of an invertible matrix, $G$, as
$\kappa_x(G)\equiv \|G\|_x \|G^{-1}\|_x$, where $x=1,2,$ or $\infty$
\cite{GVL,Higham}. Here $\|G\|_1$ is the maximum column sum norm
and $\|G\|_\infty$ is the maximum row sum norm.

\begin{thm}\label{OrderThm}
Let $(A, B)$ be a stable TIN pair ($d=1$).
If the eigenvalue magnitudes are in ascending order,
$|\lambda_{k+1}| \ge |\lambda_k|$, then $ |(M^{-1})_{i,j}| < 1$ for $i>j$
and $\kappa_2(M^{-1}) \le 2n$.
\end{thm}
\Prf
The LR factorization is 
$\left( B| A\right) =\left( M^{-1}\right) \left({e}_{1} | N\right) $. 
Now $M$ is unit lower triangular with $(M)_{k+1,k}=\lambda_k^*$. So
\BEQ
 (M^{-1})_{i,j}= (-1)^{i-j}\prod_{j\le k< i} \lambda^*_k \mu_k =
(-1)^{i-j}
        \left( \frac{1-|\lambda_i|^2 }{ 1-|\lambda_j|^2}\right)^{1/2}
        \prod_{j\le k< i} \lambda^*_k \ ,
\NEQ
 for $i>j$. Note $\|M^{-1}\|_{x} \le n $ and $\|M \|_{x} \le 2 $
for $x =1$ and $x=\infty$. Thus
 $\kappa_2(M^{-1}) \le  \kappa_1(M^{-1}) \kappa_{\infty}(M^{-1}) \le 2n$.
\eopp
This implies that the LR factorization of $(B|\ A)$
is numerically stable without pivoting.

{\em B. Eigenvectors}

We now evaluate the eigenvectors of the single input TIN system
using the parameterization (\ref{Mbidiag})-(\ref{Nbidiag})
for the case of distinct eigenvalues.
It is well known
that the eigenvectors of triangular matrices satisfy a recursion formula.
Let $V_{ij}$ be the $i$-th component of the $j$-th eigenvector
with $V_{ij}=0$ for $i<j$.
Write $V =[ V_{ij}]$ and $\Lambda \equiv $ diag
$\left( \lambda_1, \ldots \lambda_n\right)$. The eigenvector equation
$AV = V\Lambda$ becomes $NV= MV \Lambda$
or element by element:
\begin{equation}
\lambda_{k}\left( \lambda_{j-1}^*\mu_{j-1}V_{j-1,k}+V_{jk}\right) =
\left(\mu_{j-1}V_{j-1,k}+\lambda_{j}V_{jk}\right) ,
\end{equation}
We set $V_{kk}=1$, $V_{jk}=0$ for $j<k$ and solve the recursion for $V_{jk}$ when $j>k$:
\BEQ
V_{jk} =   
\frac{
\left( 1-\left| \lambda_{j}\right| ^{2} \right)^{1/2}
\left( 1-\left| \lambda_{k}\right|^{2}\right)^{1/2}
}{
\lambda_{k}-\lambda _{j}
} \
\left[
\prod_{k<j'<j}
\left(
 \frac{1-\lambda _{k}\lambda_{j'}^*}{\lambda _{k}-\lambda _{j'} }
\right) \right] \ \ ,
\NEQ
where the bracketed term is equal to $1$ for $j=k+1$.

In the next section, we show that real single input IN pairs
with complex conjugate eigenvalues
have a bidiagonal fraction representation as well. However,
we do not give an explicit parameterization of $M$ and $N$ in terms
of the eigenvalues.

\section{MIMO BAND FRACTION REPRESENTATIONS OF TIN PAIRS.}
\label{BandFracSectM}

We now prove that generic TIN pairs have a
band fraction representation: $A=M^{-1}N$, $B=M^{-1}\hat{B}$,
where $M$ and $N$ are triangular matrices of lower bandwidth $d$
and $\hat{B}$ is upper triangular.
The banded matrix fraction structure allows state space updates,
 $z_{t+1} \leftarrow Az_t + B \epsilon_t$,
in $(2d+1)n$ multiplications using (\ref{MNSAdv}).
Theorem \ref{TINpar1} and Theorem \ref{TINpar2} show how to parameterize
TIN input pairs using $nd$ parameters.

\begin{thm} \label{BandTIN1} Let $(A,B)$ satisfy $D -ADA^* =BB^*$,
where $D$ is a diagonal, positive
definite matrix and $A$ is LT and $B$ is an $n \times d$ matrix. 
Let $(B | A)$ have nonvanishing principal subminors,
$(B|A)_{1:k,1:k}$ for $k < n$, then $(B, A)$ has an
unique band fraction representation: $(B,A)=M^{-1}(\Bh,N)$,
where $M$ and $N$ are
$n \times n$ LT matrices of bandwidth $d$
and $M_{i,i}= 1$ and $\Bh_{j,k}=0$ for $j>k$.
\end{thm}

{\em Proof:} 
By Theorem 3.2.1 of \cite{GVL}, $(B | A)=L^{-1}R$ has a unique
$(B | A_{1:n,1:(n-d)})=L^{-1}R$ has a unique  
representation where R is UT and $L$ is LT with $L_{i,i}=1$.
Let $\tilde{R}$ be
the submatrix of $R$ containing columns $(d+1)$ through $(n+d)$.
Since A is LT, $\tilde{R}=LA$ is  LT of bandwidth $d$.
Note $ R( \Ib_{d} \oplus D)R^* = LDL^*$.  By Theorem 4.3.1 of \cite{GVL},
$L$ has  bandwidth $d$. We set $M=L$ and $N= \tilde{R}$
\eopp

For $d=1$, every controllable TIN pair has nonvanishing principal minors.
The condition that $(B | A)$ have nonvanishing principal minors for $k<n$
is generically true.
If $(B |A) $ has an LR decomposition and a vanishing principal minor for $k<n$,
then the induced TIN pair, $(A, B)$, does not have a unique LR decomposition
of $(B|A)$ with $M_{i,i}=1$. 
From $MM^* =NN^* +\Bh\Bh^*$,
the singular values of $M$ are singular values of $(\Bh | N)$.
When the condition that $M_{i,i}=1$ is relaxed, other band fraction
representations may be generated by $ (\Bh,N)\leftarrow D'(\Bh,N)$,
$M \leftarrow D'M$, where $D'$ is a nonsingular diagonal matrix.


The band fraction structure implies that
the $n \times (n+d)$ matrix $Y \equiv (\Bh | N)$
is upper triangular with upper bandwidth $d$.
We now examine parameterizations of the band fraction representation.
We define

\begin{defn}
Let ${\cal Y}$ denote the set of all upper triangular
 $n \times (n+d)$ matrices with  upper bandwidth $d$.
Let ${\cal Y}_1$ be the set of $Y \in {\cal Y}$ such that $Y_{i,i}=1$
for $i<n$ and $\| Y_{n,:} \|=1$ with a positive first nonzero element
in the $n$th row.
\end{defn}

Note ${\cal Y}$ has $n(d+1)$ nonzero coordinates while ${\cal Y}_1$ is
an  $nd$ dimensional manifold. We now show that we can construct a TIN pair
from any $Y$ in ${\cal Y}$ using the LQ decomposition.

\begin{thm}\label{TINpar1}
Let $Y \in {\cal Y}$ have rank$(n)$ and define $M$ and $\Qh$ as the
LQ decomposition of $Y$. Define $B =\Qh_{:,1:d}$ and $A=\Qh_{:,(d+1):(n+d)}$.
Then $M$ has lower bandwidth $d$ and is invertible and $(A,B)$ is a TIN pair.
\end{thm}

\Prf
Note rank$(M)=$rank$(Y)=n$.
Let $(\Bh | N) = Y$. Note $MM^* = YY^* = NN^* + \Bh \Bh^* $.
By Theorem 4.3.1 of \cite{GVL}, $M$ has bandwidth $d$.
From $A=M^{-1}N$, $A$ is triangular and $AA^* + BB^* = \Ib_n$
implies $(A,B)$ is a TIN pair.
\eopp


We now examine the map $f(Y) = ( A=\Qh_{:,(d+1):(n+d)}, B =\Qh_{:,1:d})$,
where $M_{ii}(Y)>0$.
From Theorem 9.1 of \cite{Higham}, $(B |A) =M^{-1} Y$ has
nonzero principal minors for $k<n$ if and only if $Y$ does.
We can rescale $Y$  and $M$ by a nonsingular diagonal matrix $D$: $Y \leftarrow DY$
and  $M \leftarrow DM$, and preserve the induced TIN pair, $(A,B)$.
This freedom and complex phase equivalence allows us to restrict
our consideration to $Y \in {\cal{Y}}_1$:

\begin{thm}\label{TINpar2}
For each TIN pair, $(A,B)$, with nonvanishing principal minors in $(B|A)$
for $k<n$,
there exists a unique $Y \in {\cal Y}_1$ and a unique diagonal
unitary matrix $E$ such that $(EAE^*,EB)$ is generated by $Y$:
 $f(Y) = (EAE^*,EB)$.
\end{thm}

\Prf
Let $(\tilde{B}, \tilde{N}, \tilde{M})$ be the unique band fraction
decomposition of $(B|A)$ with $\tilde{M}_{ii}=1$. For an arbitrary
diagonal unitary matrix $E$, the set of band fraction representations
of $(EAE^*,EB)$ is $\{(DE\tilde{B}, DE\tilde{N}E^*, DE\tilde{M}E^*) | \}$,
where $D$ is nonsingular and diagonal.
Suppose both $Y_1$ and $Y_2 \in {\cal Y}_1$ generate an equivalent
version of $(A,B)$, i.e.\  $f(Y_i) = (E_iAE_i^*,E_iB)$.
Thus $Y_i= ( D_iE_i\tilde{B} | D_iE_i\tilde{N}E_i^*)$ for some $D_i$ and $E_i$
where $D_i$ is nonsingular and diagonal and $E_i$ is diagonal and unitary.
Since $M_i = D_iE_i\tilde{M}E_i^*$,
the condition $M_{i,i}>0$ shows that $D_1$ and $D_2$ are positive matrices.
The requirement that $Y_i \in{\cal Y}_1$ forces  $|D_1| = |D_2|$ and $E_1=E_2$.
\eopp


We can parameterize input normal systems  using $Y \in {\cal Y}_1$ for $d>1$.
There are two difficulties with this parameterization.
First, the representations $(M^{-1}N, M^{-1}\Bh)$ for $(\hat{B}|N)$
in ${\cal Y}_1$
contain many equivalent TIN representations corresponding to
different orderings of the eigenvalues of $A$.
Second, in  the ${\cal Y}_1$ parameterization,
the eigenvalues of $A$ are only known after one solves $M$ by computing
the LQ decomposition of $Y$.


\section{NUMERICS} \label{NumSect}

We now discuss the computational speed of various system representations.

The bidiagonal band fraction representation, $A=M^{-1}N$, requires
only $3n$ multiplications for a state vector update
in a single input system. First, we compute
$r_t \equiv N z_t + e_1 \epsilon_t$ in $2n$ multiplications.
We then solve the bidiagonal matrix equation $M z_{t+1} =r_t$ in $n$
multiplications since $M_{i,i}=1$. 
multiplications. The matrix operations in the band fraction advance
may be implemented using the {\em dtbmv} and {\em dtbtrs} routines
from the optimized LAPACK software \cite{LAPACK}.
We caution that if $A$ has nonreal eigenvalues, then $M$ and $N$
are complex. Thus the computational advantage of the bidiagonal matrix
fraction representation is primarily limited to the case when
the eigenvalues of $A$ are real.

In \cite{HVB}, Heuberger et al. propose a  cascade realization
of IN filters. This cascade representation requires
$4n$ multiplications for a
state vector update in a single input system. 
A different cascade realization of SISO IN pairs is given in Figure 1
of \cite{NG}. The Ninness and Gustafson representation uses a lossless
cascade followed by a set of AR(1) models . This realization
requires at least $5n$ multiplications for a state update.

The direct form representations \cite{MRbook}
have even faster state advances,
but these representations are not IN and can have very
ill-conditioned Grammians \cite{MR4,Ninness}.
We advise against
using high order models that are not in input normal form.
The tridiagonal form \cite{McKelvey} is another fast representation
that is not input normal.

In \cite{MRbook, VeenVib}, embedded lossless representations are
constructed.
For $d=1$, these embedded lossless filters
$8n$ multiplications per advance \cite{MRbook}.
These embedded filters include the cost of evaluating
$C \cdot z_t$ while the IN filter representation has an additional
cost of $n$ multiplications to compute $C \cdot z_t$, so we should credit
the embedded filters with $n$ multiplications, yielding $7n$ multiplications.
Unfortunately, embedded lossless systems are not input normal.

In \cite{MR3}, we give a different representation of Hessenberg
and triangular input pairs by treating $(B,A)$ as a projection of
a product of $nd$ Givens rotations.
The multiplication count for these Givens product
representations is  $4dn$, which for large $d$ is twice as
large as the band fraction representations.
If Householder transformations are used,
the multiplication count is again $2nd$ asymptotically.
The Givens product representations are a MIMO generalization of
the cascade architecture.
This  approach may be more natural for multivariate case.
In the $d=1$ case,
the band fraction representation
has the advantage that the eigenvalues of $A$ are parameters of
the system. This eigenvalue parameterization is convenient when
the eigenvalues are prespecified or adaptively estimated.

The Givens product representations in \cite{MR3} may be less sensitive
to roundoff error since they  are orthogonal matrices \cite{MRbook}.
The computational advantage of the band fraction representation is
that fast numerical algorithms 
for band matrix inversion and multiplication are readily available.


\BEC
\section{FAST SYSTEM IDENTIFICATION}
\label{SysIdentSect}
\NEC

The band fraction representations may be used for the rapid identification
of impulse responses. We remain in the framework where a fixed basis of
orthogonal basis functions are given as described
in \cite{MRbook,Wahlberg,NG,VHB,NHG,Ninness}
and summarized in the introduction.
We then use the band fraction representation
of Sections \ref{BandFracSect}-\ref{BandFracSectM},
so $(A,B)\equiv M^{-1}(N,\Bh)$ is given.
The state vector evolves according to (\ref{MNSAdv}).
We effectively compute the second moment matrices,
$\hat{P}_t^{\delta}\equiv \sum_{i=1}^t \delta^{t-i}{z}_i{z}_i^{*}$ 
and $\hat{d}_t^{\delta}=\sum_{i=i}^t \delta^{t-1} {z}_i {y}_i^*$,
where $\delta$ is the forgetting factor.
The unknown coefficients, $\hat{C}\,$, 
using recursive least squares (RLS). 
At each time step, we reestimate $\hat{C}$ by solving
$\hat{P}_t^{\delta} {\hat{C}}_t=\hat{d}^{\delta}_t$.
This is the normal equations for the least squares estimate
 of $\hat{C}_t$. 
The $k$th component of $z_t$  represents 
 the $k$th orthonormal transfer function, ${\cal B}_k(q)$, applied to the sequence
$\{ u_1,u_2,\ldots u_t\}$. The resulting estimate of the the $j$th lead of the
impulse response is $\hat{C}A^{j-1}B$.
To solve for $\hat{C}_t$., we recommend using a QR update of the square root
of normal equations \cite{SK}.
Even faster methods are available that use the displacement rank structure
of $\hat{P}_t^{\delta}$ \cite{MR1, SK}.
 The interested reader can consult \cite{LS,Ninness}
for a comprehensive description of adaptive estimation.

The input normal filter representations are advantageous for many reasons:
First, $\hat{P}_t^{\delta} \stackrel{t\rightarrow\infty}{\longrightarrow}$
constant $\times\ \Ib_n$,  when the `true'' state space model is used
and the forcing noise is white.
Thus the regression for $\hat{C}_t$ 
is well-conditioned.
Similarly, IN filter structures are resistant to roundoff error \cite{MRbook}.
Thus IN  representations will time asymptotically satisfy
 the ansatz need by least mean squares (LMS) identification algorithms .
This leads to a second advantage of IN filters: Gradient algorithms such
as the least mean squares (LMS) algorithm often perform well enough in certain
applications to obviate the need for more complicated
and computationally intensive RLS algorithms.
Third, an $\QTR{cal}{O}(n)$ update of 
$\hat{C}_t$ is possible \cite{MR1}.
Finally, when the advance matrix is triangular,  the IN pairs form
nested families: $(A_{1:k,1:k}, B_{1:k,:})$ is TIN for $k \le n$
when $(A,B)$ is TIN. This nesting may be used in adaptive order selection.

These are advantages of all TIN filter representations.
For single input pairs,
The advantages of the band fraction representation over the cascade
representation are  primarily numerical,
the faster computational speed
($3n$ multiplications versus $4n$ or more) and the availability
of fast linear  algebra software for banded matrices.

\section{SUMMARY} \label{SummSect}
We have shown that TIN pairs generically have a matrix fraction representation,
 $A=M^{-1}N$, where $M$ and $N$ have lower bandwidth $d$.
This structure allows fast state space updates,
 $Mz_{t+1} \leftarrow Nz_t+ \hat{B} \epsilon_t$,
and fast solution of matrix equations: $Ax = f$.
The total operations count is $(2d+1)n$ multiplications.
We believe that our work is the first study of system representations,
where $A$ is a matrix fraction of banded matrices, $A=M^{-1}N$.


For single input filter design,
our representations are parameterized explicitly
in terms of the eigenvalues, $\{\lambda_k\}$. 
When $M$ is bidiagonal, solving $Mz_{t+1} =f_t$ requires only ${\cal{O}}(n)$
multiplications and may be implemented as a systolic array with
order independent latency. As showed in Theorem \ref{OrderThm},
if the eigenvalues are ordered in ascending magnitude,
then the matrix inversion is well-conditioned and
suitable for fixed point operations.
For system identification, 
the covariance of $z_t$ tends to the identity matrix
when the forcing noise is white.
Thus least squares regression
is both well-conditioned and computationally fast.
Since $A$ is triangular, one can compute the evolution of  all embedded models
of dimension $n'< n$ simply by projecting the model of dimension $n$ onto the first
$n'$ coordinates. Thus the triangular structure simplifies the evaluation 
of this nested family of models for model order selection.



\end{document}